%% file: preprint_Sergeev_CD_actuator_as_sound_absorber.tex
\documentclass[onecolumn]{article}

\usepackage[modulo,switch]{lineno}
\modulolinenumbers[1]

\usepackage[utf8]{inputenc}
\usepackage{amsmath}
\usepackage{amssymb}
\usepackage{graphicx}
\usepackage{cite}
\usepackage{color,soul}
\usepackage{dblfloatfix}    
\newcommand{\figref}[1]{\mbox{Figure~\ref{#1}}}
\newcommand{\vect}[1]{\boldsymbol{#1}}

\makeatletter\@ifundefined{date}{}{\date{}}
\makeatother

\evensidemargin\oddsidemargin \hoffset-22.4mm \voffset-28.4mm
\begin{document}

\title{Corona Discharge Actuator as an Active Sound Absorber Under Normal and Oblique Incidence\footnote{This work has been submitted to Acta Acustica for possible publication. Copyright may be transferred without notice, after which this version may no longer be accessible.}
}

\author{Stanislav Sergeev$^{1)}$, Thomas Humbert$^{2)}$, Hervé Lissek$^{1)}$, Yves Aurégan$^{2)}$ \\
$^{1)}$ Signal Processing Laboratory LTS2, EPFL, CH-1015 Lausanne, Switzerland.\\
$^{2)}$ Laboratoire d'Acoustique de l'Universit\'{e} du Mans,\\ LAUM-UMR CNRS 6613, Avenue Olivier Messiaen, 72085 Le Mans Cedex 9, France.}

\maketitle\thispagestyle{empty}
\begin{abstract}
In the majority of active sound absorbing systems, a conventional electrodynamic loudspeaker is used as a controlled source. However, particular situations may require an actuator that is more resistant to harsh environments, adjustable in shape, and lighter. In this work, a plasma-based electroacoustic actuator operating on the atmospheric corona discharge principle is used to achieve sound absorption in real-time. Two control strategies are introduced and tested for both normal in the impedance tube and grazing incidence in the flow duct. The performance of plasma-based active absorber is competitive with conventional passive technologies in terms of effective absorption bandwidth and low-frequency operation, however, it presents some inherent limitations that are discussed. The study reveals that the corona discharge technology is suitable for active noise control in ducts while offering flexibility in design, compactness, and versatility of the absorption frequency range.
\end{abstract}

\section{Introduction}
\input{introduction.tex}

\section{Absorption concepts and corona discharge actuator}
\input{section_2_concept}

\section{Assessment of the control strategies under normal incidence}
\input{section_3_normal}

\section{Assessment of the control strategies under grazing incidence}
\input{section_4_grazing}

\section{Conclusion}
\input{section_5_conclusion}

\section*{Acknowledgements}
This   study   has   received   funding   from   the   European Union’s  Horizon  2020  research  and  innovation  program under grant agreement No 769350.

\small
\bibliographystyle{ieeetr}

\bibliography{Biblio}

\end{document}

%% file: introduction.tex
\label{introduction}

Recent aircraft engines developments have been focusing on decreasing the fan's rotation speed, with the aim of reducing their fuel consumption. Consequently, the frequencies of the tonal noise emitted with such architectures are lower than for previous engine generations~\cite{palma2018acoustic, ma2020development}. Absorbing sound within this new frequency range would not be such a challenge if the space and the weight allocated to the acoustic treatment were not limited. Thus, conventional aircraft liners based on quarter wave-length resonators~\cite{maa1998potential,beck2015impedance} may not be an optimal solution anymore, and new sub-wavelength liner concepts have to be found.

A good example of such materials are membrane-based acoustic resonators which allow increasing the ratio between the acoustic wavelength to be absorbed and the absorber thickness (in~\cite{auregan2018ultra} the ratio reaches $~200$). However in this case, the high sub-wavelength ratio is associated with a very narrow absorption peak that is not well suited for nacelles applications.
Slow sound devices have been also investigated in the framework of this problem, but dramatic flow effects where found to be critical for their performance and final use in the aeronautic context~\cite{auregan2016low}. Finally, recent progress in modelling and 3D printing of porous and meta-porous materials placed them as serious candidate concepts to perform broadband absorption while reducing the mass of the acoustic treatments~\cite{boulvert2019optimally,cavalieri2020graded}. However, their application in real environmental conditions is still to be evaluated.

Alternatively, active sound absorbing systems have been intensively investigated. Different approaches demonstrated their ability to manipulate the sound field and to impose a certain acoustic impedance on the controlled interface in a wide frequency range\cite{olson1953electronic, thenail1994active, furstoss1997}. Active absorbers are specifically efficient at low frequencies while staying relatively small compared to the wavelength \cite{rivet2016}. Several acoustic liner prototypes operating with active control principles have already been reported \cite{galland2005, betgen2012implementation, boulandet2018}. Moreover, as it is possible to digitally change control parameters, absorption can be focused on selected frequencies, which is of high interest when considering a drifting tonal noise emitted by an aircraft engine.

 When implementing active noise reduction methods, typically electrodynamic loudspeakers are employed as control transducers. It is a favorable choice for many applications, since loudspeaker drivers have rather wide frequency response. Moreover, they are inexpensive and can be precisely modelled analytically \cite{rossi1988}. However, for certain situations, such as an aircraft nacelle, their utilization may appear difficult. In particular, the loudspeaker membranes constitute a fragile part in such extreme conditions. Additionally, a great number of active cells needed to cover large area \cite{boulandet2018} is likely to yield an excessive total system weight. Thus, piezoelectric transducers have been proposed as a substitute since they are more compact and lighter \cite{galland2005}. Unfortunately, their narrow frequency response limits their application for active control \cite{betgen2012implementation}. Therefore, an alternative actuator which does not exhibit the aforementioned shortcomings and can be used for active noise control is still to be found.

This study focuses on a plasma-based actuator as a possible solution for active control applications in aircraft engines. It is worth noting that these actuators are widely used in the flow control research \cite{moreau2007airflow, thomas2008plasma, el1997drag}. Some acoustic applications such as instability wave control in a turbulent jet and tonal cavity noise reduction have been reported in the literature \cite{kopiev2014instability, huang2008streamwise}. In general, a plasma actuator simply consists of two metallic electrodes separated by a dielectric gap. Several types of gas discharges such as dielectric barrier and corona discharge can be used to control the velocity field in the vicinity of the actuator. This is done by generating a volume force due to the constant air ionization induced by a strong electrical field. When an alternating electrical signal is applied to the plasma actuator, the modulation of the body force and the heat release in the vicinity of the electrodes lead to the perturbation of the local pressure and velocity fields. Thus, sound waves are created \cite{bastien1987, bequin2007}. The common advantage of plasma actuators is a simple construction without any moving parts. Consequently, they are light, mechanically robust, with a short response time to an electrical signal and a wide operating frequency range.

In \cite{sergeev2020}, the acoustic properties of an electroacoustic actuator based on the atmospheric corona discharge in a wire-to-mesh geometry have been investigated. The transducer was found to be a potential candidate for active noise control applications. In the present work, we further investigate such capabilities by implementing two active noise reduction strategies called "hybrid absorption" and "active impedance control". Their performance with a plasma actuator are evaluated and compared. The paper is divided in three main parts. First, the two sound absorption methods are introduced. Then, experimental assessment of two prototypes with each control strategy is carried out in an impedance tube under normal incidence. Finally, the prototype of an acoustic liner is presented and experimentally characterised under grazing incidence. Its capabilities to induce broadband sound absorption and to attenuate a certain frequency interval are presented. The advantages and shortcomings of the developed system are discussed in the conclusion.

%% file: section_2_concept.tex
\label{concepts}
Previous study \cite{sergeev2020} introduced simple analytical model of a corona discharge (CD) actuator sound radiation in free space with far field approximation. Nonetheless, it does not describe the near field behaviour when the system is enclosed. In addition, the radiated sound pressure is linked to such parameters as effective ion mobility which in turn depends on gas mixture, density and humidity variable in different environments \cite{zhang2017}. This makes the active control strategies, which rely on the physical model of the transducer, more difficult to implement. Therefore, two noise control strategies were chosen and implemented which do not require the actuator's analytical model.
The first technique, referred to as "hybrid absorption", utilizes the absorbing properties of passive structures combined with the active control of the actuator.
The second method consists in a direct active impedance control, using a pressure-velocity feedback. 

\subsection{Hybrid absorption}

When a constant pressure difference is applied between the two opposite sides of a porous layer of thickness $d$, the velocity $u$ of the generated fluid flow can be written as:
\begin{equation}\label{eq1}
u=(p_1-p_2)/rd,
\end{equation}
where $r$ is the flow resistivity of a material. The same equation holds for acoustic pressure and velocity at low frequencies assuming the porous layer is purely resistive \cite{furstoss1997, galland2005}. If the pressure behind the porous layer $p_2$ is set to zero, the material flow resistance $rd$ becomes equal to the acoustic impedance at the front surface of the layer: 
\begin{equation}\label{eq2}
rd=Z_1=p_1/u.
\end{equation}
In the hybrid absorption strategy, the CD actuator is employed to impose a null pressure $p_2=0$ behind the porous layer at each frequency. To obtain perfect absorption, the problem consists in finding a material with a resistance close to the optimal impedance. Since the latter is different for the normal and grazing incidence experiments, the implementation details are provided in the following sections. 

This approach is advantageous for the corona discharge transducer since it only relies on pressure measurements close to the CD and requires a low signal distortion from the actuator. The main drawback of this technique, however, is that there is no possibility to influence the reactive part of the acoustic impedance. Instead, it is always set to zero. More sophisticated method involving the control of both $p_1$ and $p_2$ can be used to target frequency dependent impedances, but in this case another resistive layer will be needed to protect the exposed microphone against the flow which complicates the design \cite{betgen2012implementation}.

\subsection{Direct impedance control via pressure-velocity feedback}
\label{concept_direct}

\begin{figure}[h]
 \centerline{
 \includegraphics[width = \columnwidth]{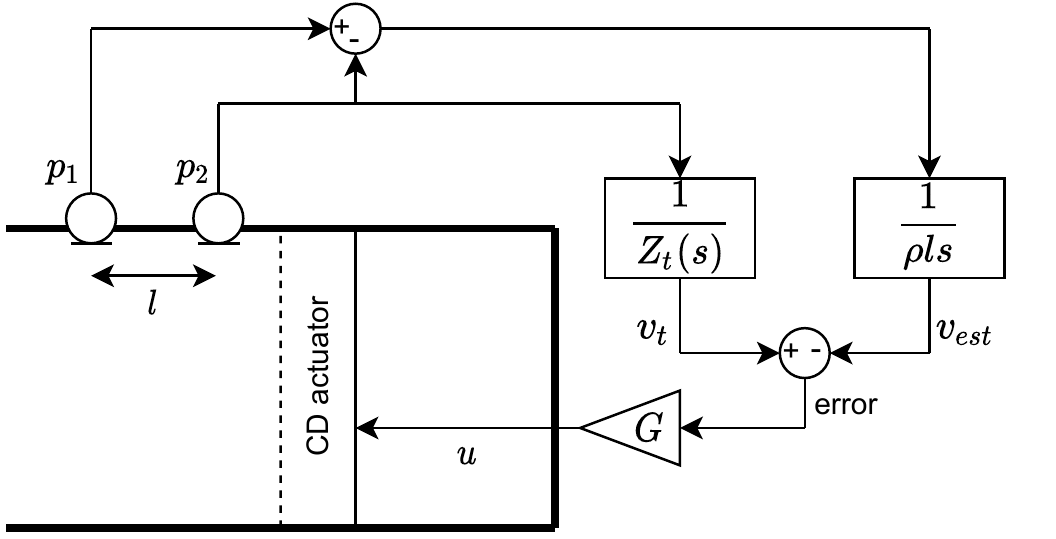}}
 \caption{Schematics of the direct impedance control loop via pressure-velocity feedback.}
 \label{fig2}
\end{figure}

If the actuator model is unknown, the measurements of acoustic pressure and velocity can be used for applying a direct active acoustic impedance control (\figref{fig2}). 

The estimation of the particle velocity $v_{est}$ in time domain using two closely spaced microphones $p_1$ and $p_2$ is obtained from the one dimensional Euler equation. Approximating the pressure gradient by the difference between pressure measurements at two positions, the equation reads:
\begin{equation}\label{eq4}
v_{est}(t) = - \frac{1}{\rho} \int \frac{\mbox{d} p(t)}{\mbox{d} x} d t \approx \frac{1}{\rho} \int \frac{p_1(t)-p_2(t)}{l} \mathrm{d}t,
\end{equation}
where $l$ is the distance between the microphones. In the frequency domain, equation (\ref{eq4}) becomes
\begin{equation}\label{eq5}
v_{est}(s) = \frac{p_1(s)-p_2(s)}{\rho l s},
\end{equation}
with $s= j \omega$ the Laplace variable and $\rho$ the air mass density. Such an estimation of the particle velocity is only valid if $l$ is reasonably smaller than the wavelength of the acoustic signal. On the other hand, when the wavelength is much larger than $l$, the separation may be not sufficient for an accurate estimation of the velocity in presence of any parasitic noise in the system. Therefore, the distance $l$ should be carefully chosen in order to provide a meaningful velocity estimation depending on the frequency range of interest.

As can be seen in \figref{fig2}, the two microphones used to estimate the particle velocity are placed close to the front face of the CD actuator. The estimated velocity $v_{est}$ is further compared to the target $v_t$. The target velocity is the one needed to achieve the target acoustic impedance $Z_t$ for a given input acoustic pressure value. Thus, it is calculated in the frequency domain by dividing the pressure from microphone 2 by the target impedance:
\begin{equation}\label{eq3}
v_t = \frac{p_2(s)}{Z_t(s)}.
\end{equation}
The steps corresponding to equations (\ref{eq5}) and (\ref{eq3}) are shown in \figref{fig2} with rectangular blocks indicating continuous time transfer functions which should be discretized for implementation on a digital platform. The difference between $v_t$ and $v_{est}$ represents the error that should be minimised. This error is multiplied by a dimensional gain $G$ to provide the appropriate voltage signal $u$ to the actuator. 

Compared to hybrid absorption where only constant real valued impedances can be achieved, and where any adjustment requires a replacement of the porous layer, the active impedance control method allows changing it digitally. It can also be set as a frequency dependent function. However, the exact value of target impedance can never be achieved since it requires infinite gain and is affected by the presence of delays in the control and physical sound propagation. Nevertheless, the actually achieved impedance tends to the target value as gain $G$ increases.

\subsection{CD actuator}

\begin{figure}[h]
 \centerline{
 \includegraphics[width = 6.5 cm]{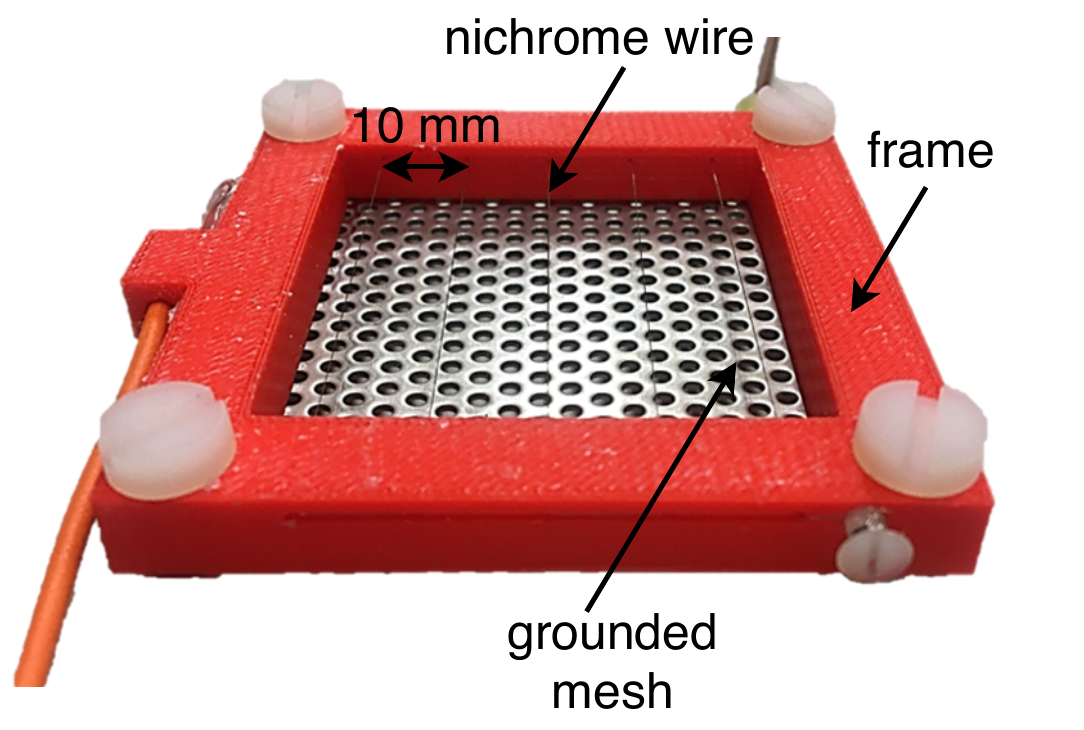}}
 \caption{Photo of corona discharge actuator in a wire-to-mesh geometry.}
 \label{fig3}
\end{figure}

The experimental prototype of the corona discharge actuator used for both control strategies has been built in a wire-to-mesh geometry (\figref{fig3}) according to the design proposed in \cite{sergeev2020}. The first electrode is made of nichrome wire with 0.1 mm diameter strung on a plastic frame, forming a pattern of thin parallel wires in one plane. A perforated stainless steel plate is used as the second electrode. The air gap between the electrodes is 6 mm. For the impedance tube measurements, the operating area of the actuator is $50 \times 50 \mbox{ mm}^2$, as it can be seen in \figref{fig3}. For the experiments under grazing incidence, the area has been increased to $50 \times 70 \mbox{ mm}^2$. When a positive 8.2 kV DC voltage is applied to the nichrome wire, the actuator provides a stable corona discharge. Adding the alternating voltage component to the constant one leads to the sound generation \cite{klein1954, bastien1987, bequin2007}. The harmonic distortion of the acoustic signal does not exceed 10 \%, which is satisfactory for the active control application \cite{sergeev2020}.

%% file: section_3_normal.tex
\label{normal}
\subsection{Experimental setup and implementation}

\begin{figure}[h]
 \centerline{
 \includegraphics[width = \columnwidth]{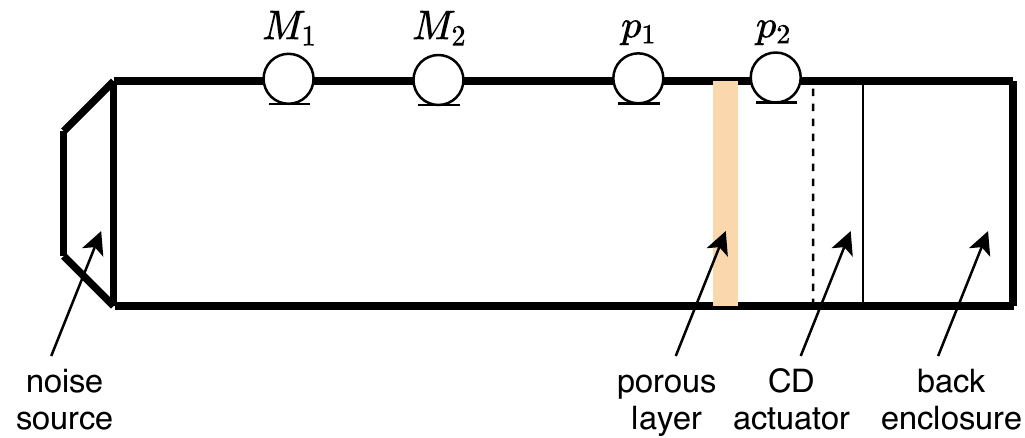}}
 \caption{Schematic of the normal incidence experiments. The porous layer is only installed for the hybrid absorption method, whereas $p_1$ is only used for the active impedance control. Not to scale.}
 \label{imp_tube}
\end{figure}

For the purpose of the normal incidence experiments, an impedance tube with a rectangular cross section of $50 \times 50 \mbox{ mm}^2$ was used (\figref{imp_tube}). A noise source represented by a loudspeaker is installed at one side of the tube. At the other end, the active absorber is mounted. The CD actuator is enclosed with a 20 mm cavity from the high voltage side. Two microphones $M_1$ and $M_2$ estimate the achieved absorption coefficient and acoustic impedance according to ISO 10534-2 standard. PCB 130D20 ICP microphones (cutoff 50 Hz) are used in the experiment for both the measurements and the control ($M_1$, $M_2$, $p_1$,  $p_2$ in \figref{imp_tube}).
Perfect absorption is obtained when the sample impedance matches the characteristic air impedance value $Z_0=\rho c$, where $c$ is the speed of sound in air. For the hybrid absorption method, a porous layer with a resistance close to this impedance is required. To that end, four layers of the same wire mesh are stacked together resulting in a total structure thickness of 1.5 mm with a resistance of $1.07\rho c$. It is installed 10 mm ahead from the actuator. 

The microphone $p_2$ located close to the front face of the actuator is used to minimize the acoustic pressure behind the porous layer. The control procedure of the hybrid absorption method is based on the adaptive least mean square algorithm (FxLMS), which takes a (buffered to $n$ previous values) reference noise signal $\vect{p}(t)$ from $M_2$ as an input, and the pressure signal $e(t)$ from $p_2$ as an error. At each time step $t$, the weights vector $\vect{W}(t)$ of the adaptive filter changes according to:

\begin{equation}\label{eq6}
\vect{W}(t) = \vect{W}(t-1) + \mu e(t)\vect{p}^T(t).
\end{equation}
In (\ref{eq6}), $\mu$ is an adaptation step size. When the weights are updated, the reference noise signal is filtered out using these coefficients, inverted, and played back to the CD actuator as the AC voltage component. The algorithm runs at the sampling frequency of $20$ kHz. The calculated output voltage from the controller for the CD actuator is reduced by a factor of 1000. Before connecting to the actuator the signal passes the high voltage amplifier (TREK 615-10) with a voltage gain of 1000.

To estimate the particle velocity, the active impedance control method takes the signals from two microphones $p_1$ and $p_2$, as described in \figref{fig2}. The distance $l$ was set to 30 mm, providing an accurate velocity estimation from 100 to 2000 Hz. Compared to the setup for hybrid absorption, the porous layer is removed, while the other aspects are similar.

\setcounter{figure}{3}
\begin{figure*}[!h]
  \centering
  \includegraphics[width = 8cm]{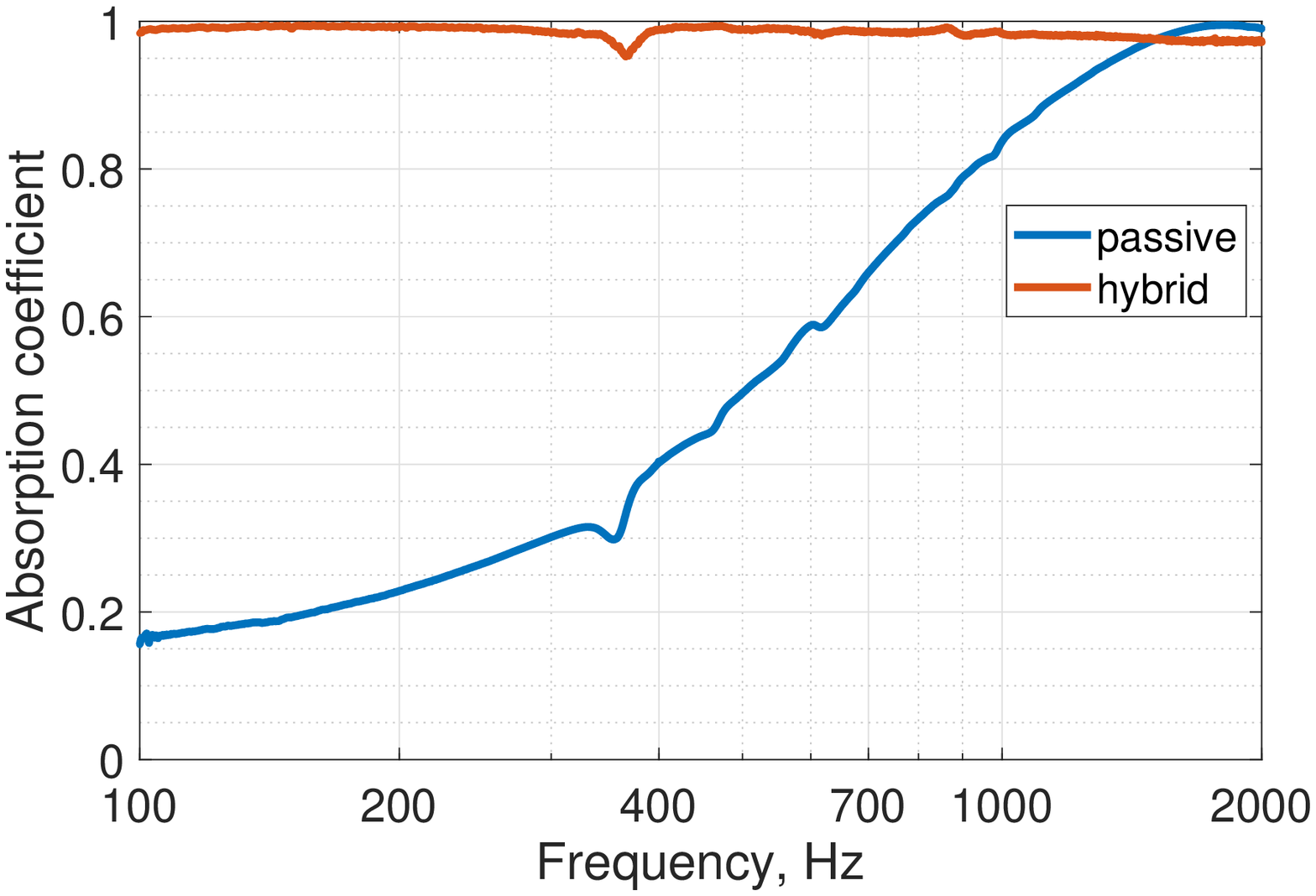}
  \includegraphics[width = 8cm]{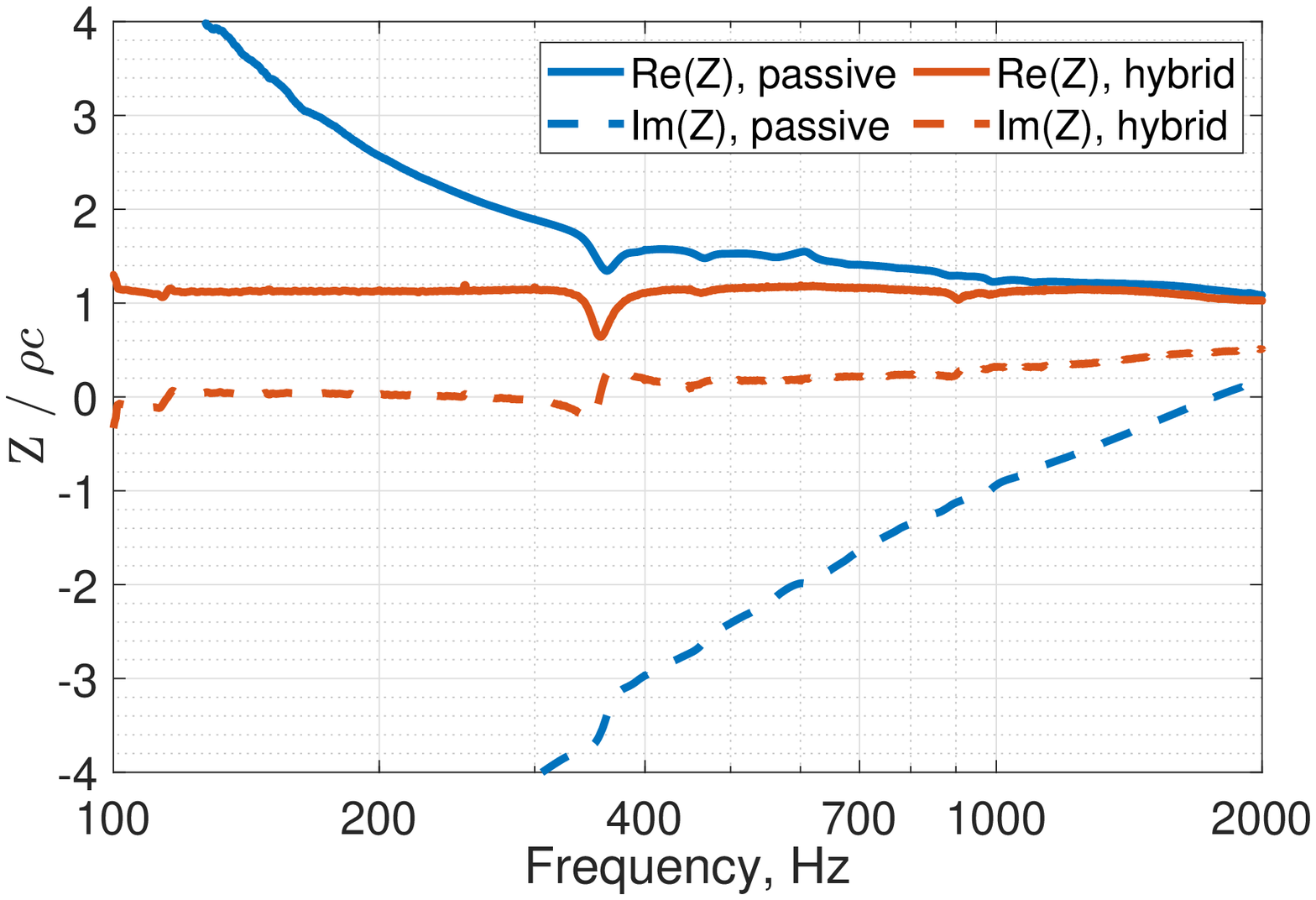}
  \caption{Hybrid absorption, normal incidence. Sound absorption coefficient (left) and acoustic impedance (right) measured  for the CD actuator when powered (red curves) or not (passive response, blue curves). A wire mesh with a resistance of 1.07$\rho c$ is installed.}
  \label{N_HA}
\end{figure*}

\subsection{Results with hybrid absorption}

\figref{N_HA} displays the measured absorption coefficients and acoustic impedance as the functions of frequency. The quantities are obtained for the passive case (CD actuator is switched off) and when the actuator is controlled with the hybrid method. In the passive regime, sound absorption is mainly due to the presence of the wire mesh in the tube. The stainless steel grounded plate from the CD actuator was measured to have a resistance of $0.02\rho c$, thus, its contribution to the total sound absorption is assumed to be negligible. Therefore, the passive termination forms a quarter wave length resonator, which reaches a peak of absorption around 1850 Hz. When the control is on, the system becomes almost a perfect absorber in the whole frequency range from $100$ to $2000$ Hz. Indeed, the absorption coefficient remains close to one at low frequencies and gradually decreases to 0.97 for the highest measured frequencies. 

The achieved acoustic impedance measured at the position of the wire mesh is depicted in \figref{N_HA} (right). The real part of the imposed impedance remains almost constant around 480 Pa$\cdot$s/m, which falls close to the measured wire mesh resistance. In the low frequency range, the reactance stays close to zero. However, a steady rising trend can be observed above 500 Hz, suggesting the presence of an acoustic mass in the porous layer. Therefore, the assumption of a purely resistive material is not completely valid anymore. At frequencies higher than $2$ kHz, we thus expect a further increase of the controlled reactance, leading to a degradation of the absorption performance. Nevertheless, the hybrid absorption method applied to the CD actuator permits to successfully change the interface acoustic impedance from high magnitudes in both resistance and reactance towards lower target values, especially in the low frequency range. This leads to broadband sound absorption in the impedance tube.

\begin{figure*}[!t]
  \centering
  \includegraphics[width = 8cm]{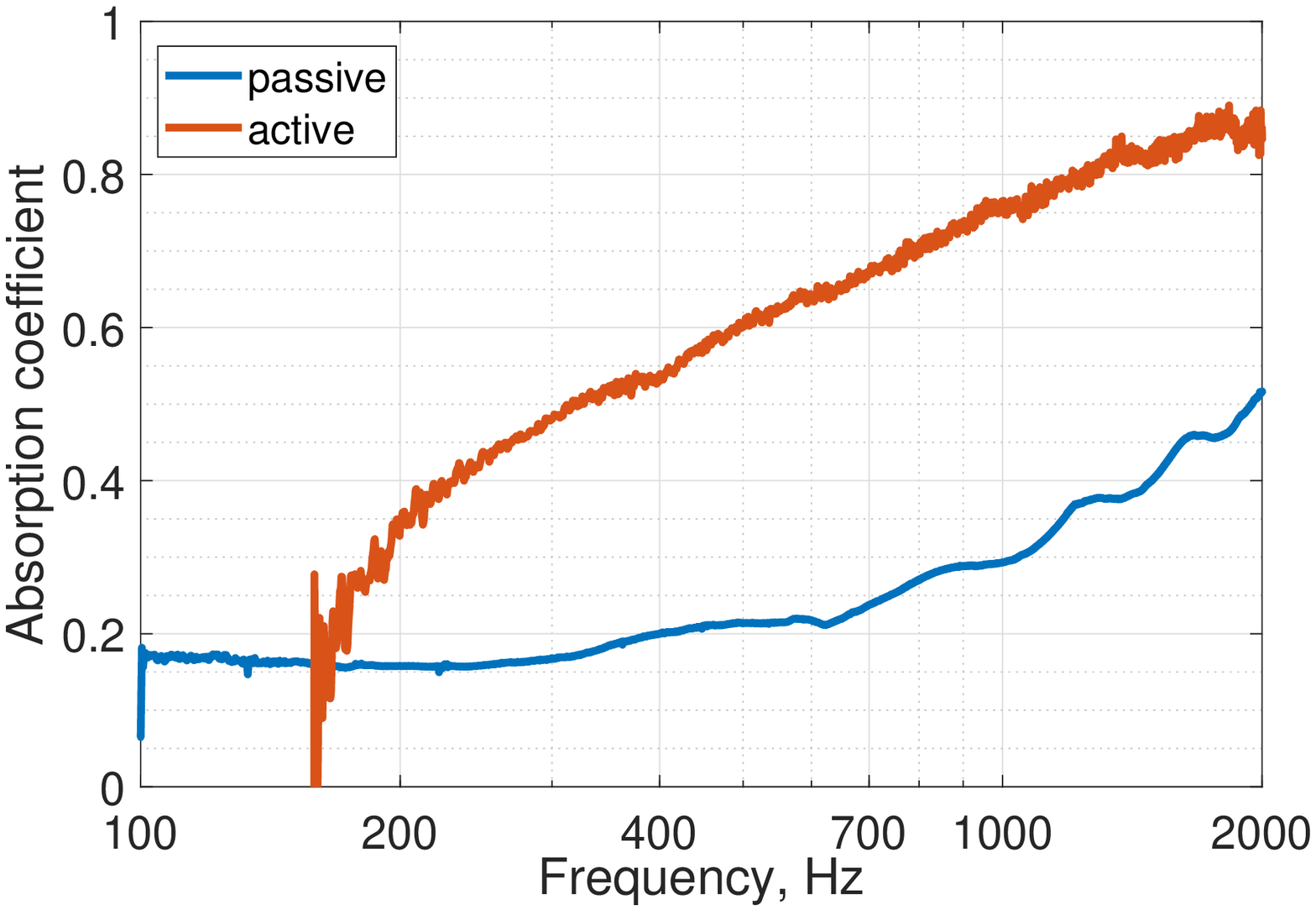}
  \includegraphics[width = 8cm]{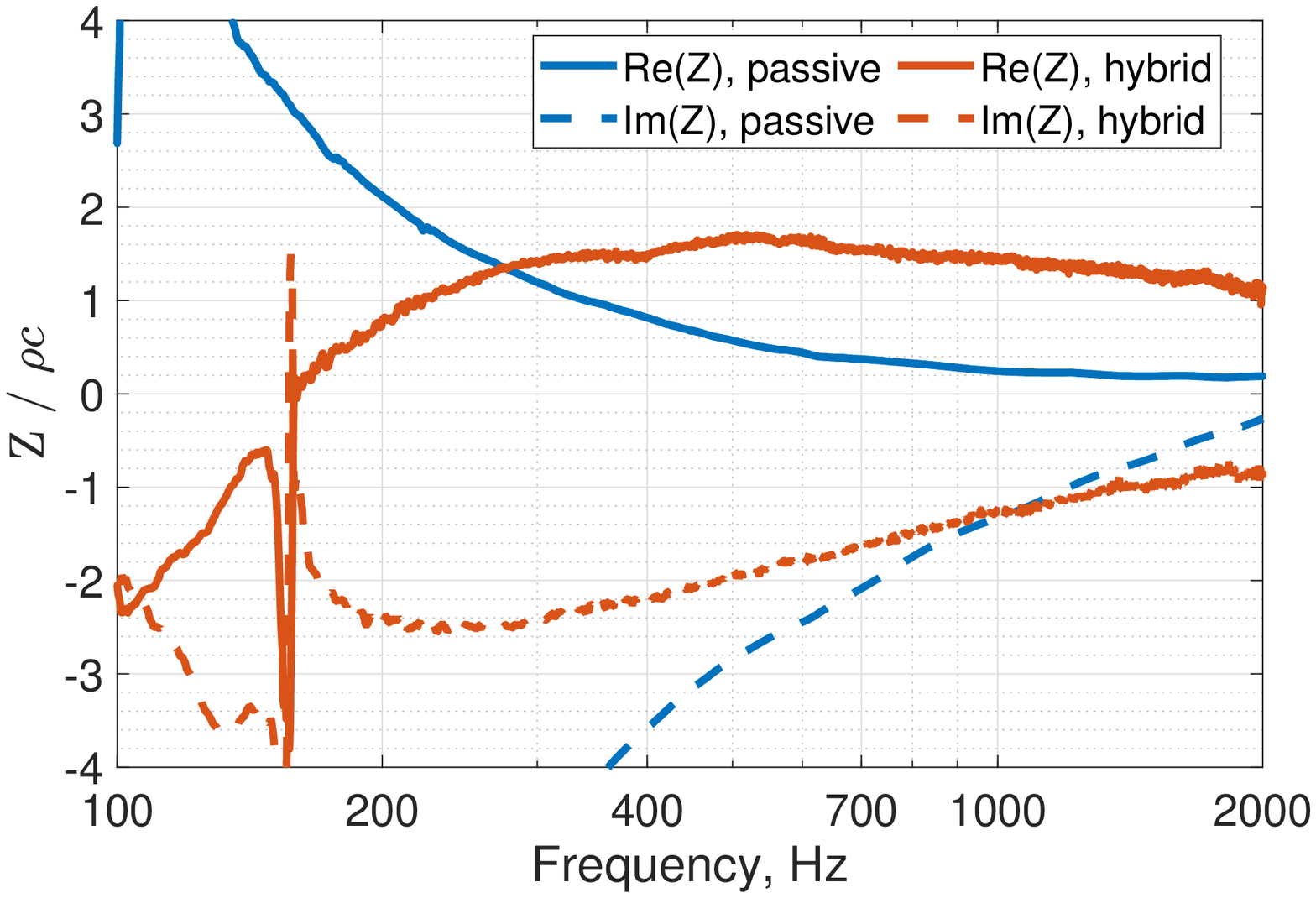}
  \caption{Active impedance control, normal incidence. Sound absorption coefficient (left) and acoustic impedance (right) measured at the front face of the the CD actuator when it is powered (red curves) or not (passive response, blue curves).}
  \label{N_DIC}
\end{figure*}

\subsection{Results with direct impedance control}

The direct impedance control method was also tested in the impedance tube. To that end, the porous layer is removed and a 10 mm layer of melamine foam has been mounted on the enclosure wall in order to increase the algorithm stability at high frequencies. As experiments were carried out in the tube under normal incidence, the target impedance was set to the specific impedance of air  $\rho c \approx 410$ Pa$\cdot$s/m. 

The achieved absorption and impedance are displayed in \figref{N_DIC} for both passive and active systems. In the passive case, the real part of the measured impedance is significantly lower than the target impedance above 400 Hz (blue line in the right panel of \figref{N_DIC}), whereas the increasing imaginary part corresponds to the cavity compliance. The resulting absorption coefficient remains within the range $0.15$ – $0.2$ at low frequencies. A further growth up to 0.5 for higher frequencies is caused by the presence of the thin absorbing melamine layer at the termination. When the CD actuator is controlled via the feedback loop, the impedance changes considerably. In particular, the real part (red line in the right panel of \figref{N_DIC}) shifts towards the target value of $410$ Pa$\cdot$s/m. Although the target value is purely real, the achieved reactance is not equal to zero because of the limited maximal gain and unavoidable delays. Note that the active measurement presented here corresponds to the maximal gain value $G$ at which the control remains stable. For higher gains, the system tends to instability, and ways to address this issue need to be further investigated in order to achieve better performance.

In the implemented feedback scheme the final voltage applied to CD actuator is proportional to the velocity difference (\figref{fig2}). The particle velocity estimation contains the integration step, which can accumulate a constant or non periodically changing signal and lead to the presence of significant voltage offset in the output. Finally, such an error can move the DC voltage powering the actuator out of the corona discharge region, leading to either too low voltage for air ionization or too high voltage transmitting discharge to an electrical arc. To avoid this, the high pass filtering of the input signals from $p_1$ and $p_2$ microphones is added which leads to the ineffectiveness of the active system below 170 Hz.

Despite abovementioned limitations at low frequencies, the impedance achieved by the direct impedance control technique induces a substantial increase in sound absorption in the whole frequency range, compared to the passive case.

%% file: section_4_grazing.tex
\label{grazing}

\begin{figure*}[!b]
  \centering
  \includegraphics[width=17 cm]{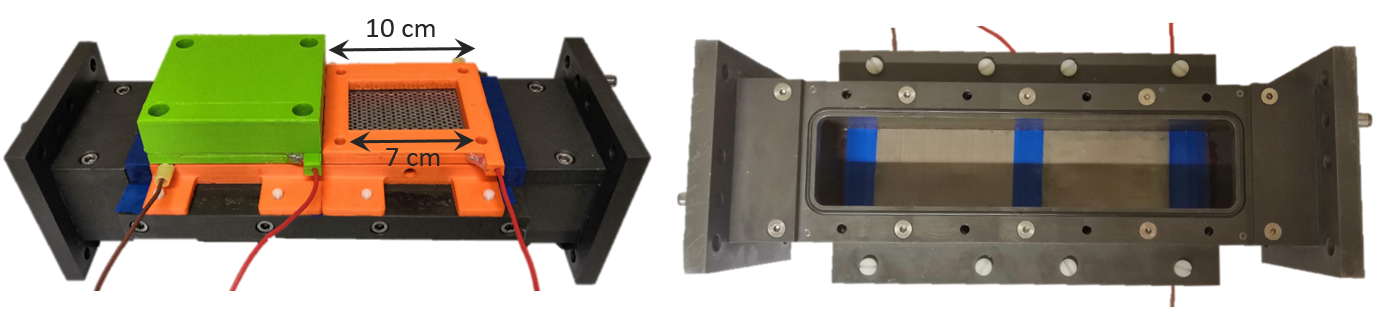}
  \caption{Test section for grazing incidence experiments with two mounted active cells. Top view on the left (one cell is partially disassembled),  bottom view with removed duct wall on the right.}
  \label{CD_liner}
\end{figure*}

\subsection{Experimental setup}
Noise reduction in a duct under grazing incidence requires an adjustment of the control parameters, as well as a slight modification of the actuator geometry.

In the impedance tube, below the first cutoff frequency the actuator size makes no difference for the control, since the wave fronts are parallel to the actuator surface. This means that the acoustic pressure at any point of the actuator plane is the same. However, it is not the case for grazing incidence, where the sound waves propagate along the actuator plane. Such configuration imposes a restriction on the actuator size: the length of the active cell should not be larger than half of the acoustic wavelength of interest. In this study, the setup includes 2 active cells (\figref{CD_liner}). Each one covers an area of $10\times5$ cm$^2$, although the operating area is restricted to $7\times5$ cm$^2$. With such dimensions active control can be efficient up to approximately 2 kHz. To tackle higher frequencies, more cells should be arranged per unit length, and the control delays should be reduced. 

Each cell comprises one or two microphones depending on the control strategy. The input signals are processed separately, and the individual output signals are given to the actuators through two high voltage amplifiers. The actuators are designed in a modular way. Two cells are screwed side by side to the duct wall. First, the cavity for the microphone is screwed to the duct. Then, the perforated plate (ground electrode for corona discharge) is placed on top. Further, the plastic frame with high voltage wires is positioned. Finally, the system is enclosed with a rigid plastic wall set to 2 cm, as in the normal incidence case.

\figref{CD_liner} illustrates the mounting of the actuators for the hybrid absorption method. For the direct impedance control, a cavity hosting two microphones should be installed. Since a certain minimal distance should be kept between the sensors in order to resolve the low frequency components, the system is more cumbersome than for hybrid absorption.

\subsection{Flow duct}
\label{sec:section_CNRS}

\begin{figure*}[h]
 \centering
 \includegraphics[width = 8cm]{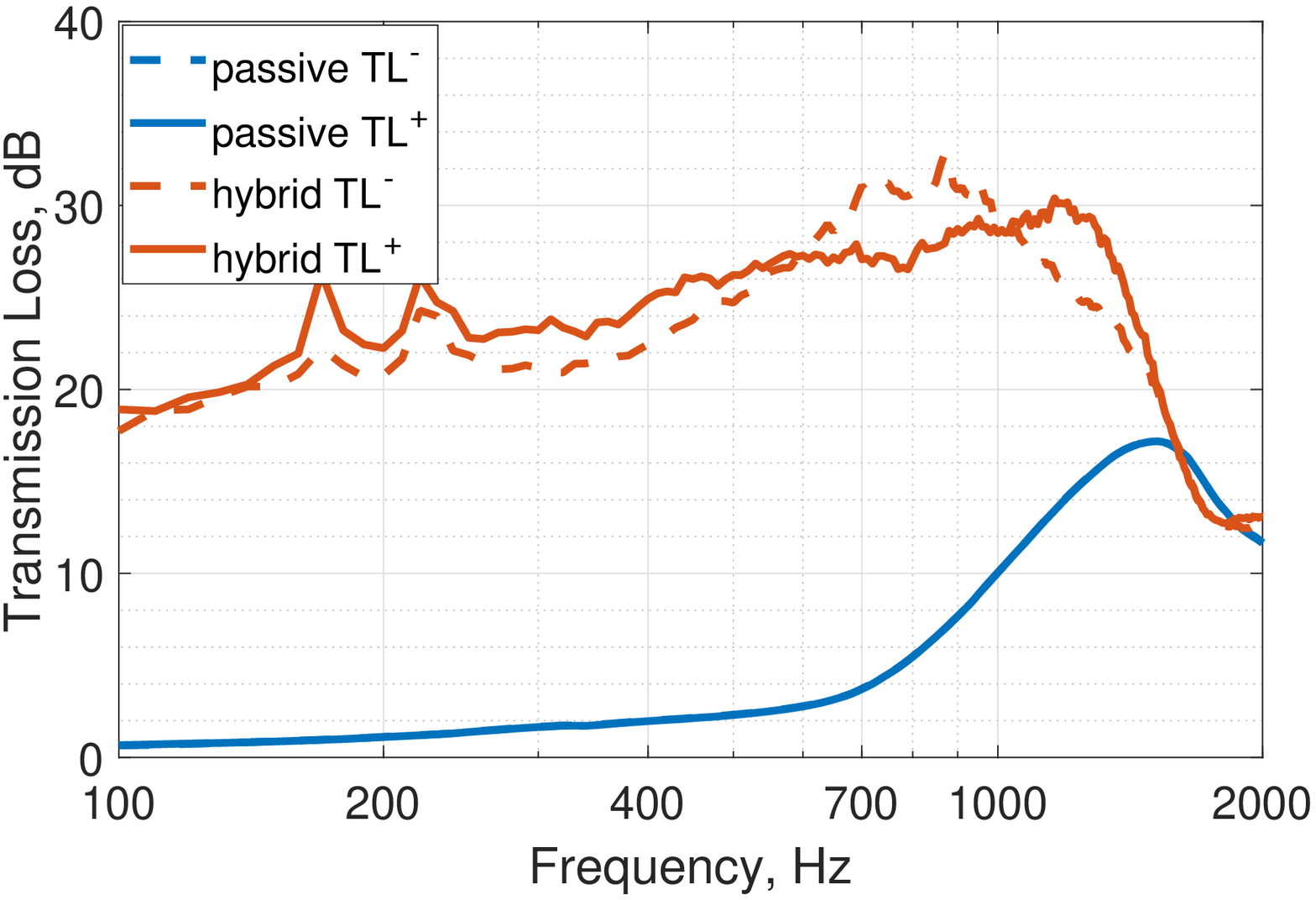}
  \includegraphics[width = 8cm]{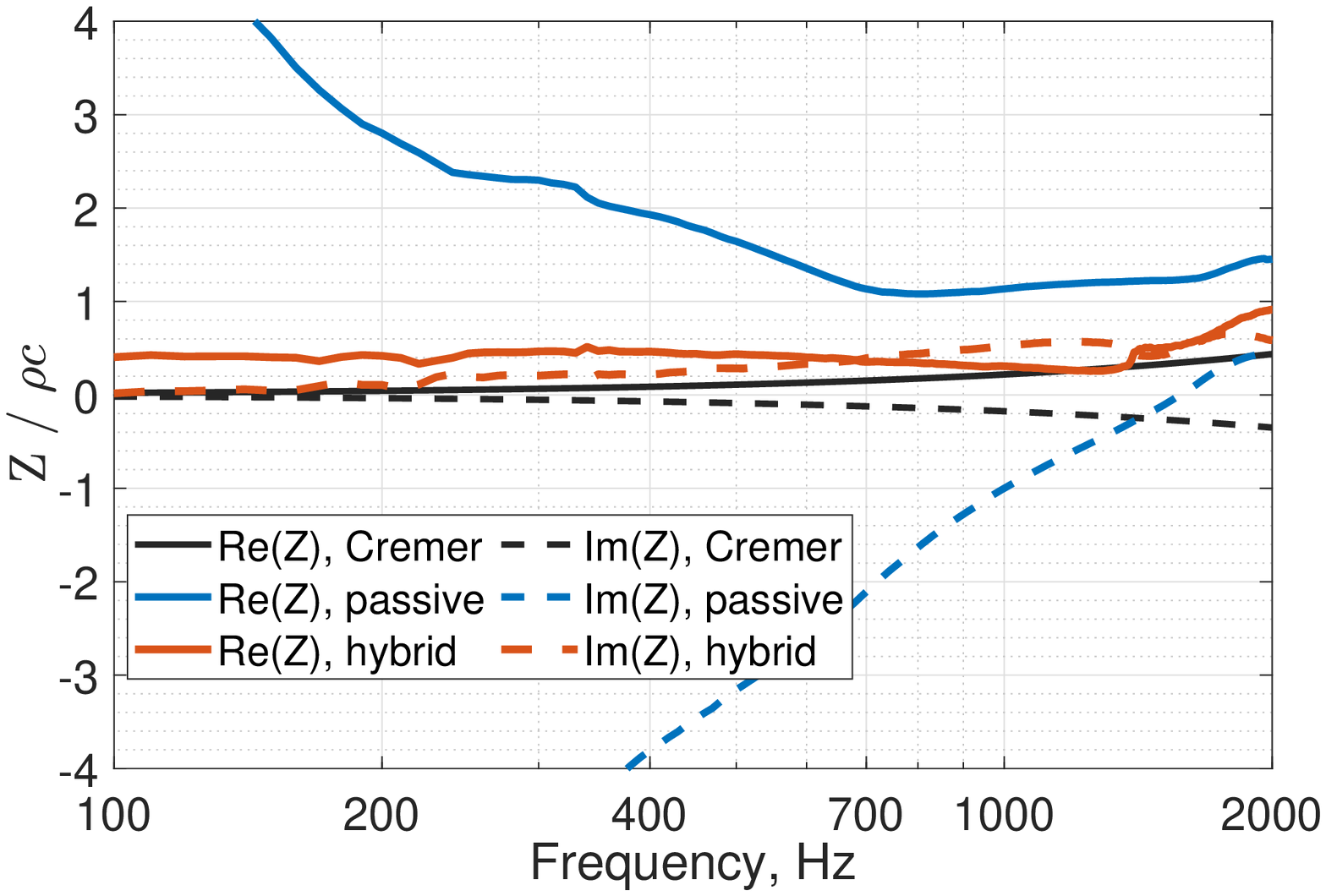}
 \caption{Transmission loss (left) and normalized acoustic impedance (right) of hybrid absorption system in comparison to passive behaviour without mean flow. The Cremer impedance is calculated for a duct height of 40 mm.}
 
 \label{H_M0}
\end{figure*}

The performance of the plasma-based acoustic liner was evaluated in the rectangular duct with cross section of $ 40 \times 50$mm$^2$. It is plugged to a fan on one side, and ended by an anechoic termination on the other side. The acoustic characterisation of the actuators arrangement is performed using six flush-mounted microphones, with three of them placed on each side of the measurement section. Acoustic plane waves are generated by a compression chamber which is mounted either upstream or downstream relative to the test section. The signal consists in a sinusoidal sine sweep going from $100$ to $2000$ Hz with steps of $10$ Hz. A constant amplitude for the incident acoustic wave is maintained on the whole frequency range. 

The scattering matrix of the corona discharge actuator is computed using two different acoustic states: one with the upstream source on (downstream source off) and one with the downstream source on (upstream source off). At each frequency step, the acoustic pressure on each microphone is obtained by averaging over 400 cycles without flow, and over 1000 cycles with flow. The scattering matrix consists of the reflection coefficients ($R^+$,$R^-$) and of the transmission coefficients ($T^+$,$T^-$). The $+$ symbols correspond to the case where the incident plane waves come from upstream the sample. The $-$ symbols indicate that the waves come from downstream. Then, the performance of the plasma-based liner is evaluated in terms of transmission losses defined as:
\begin{equation}\label{eq7}
    TL^\pm= 20\mbox{log}10\left|\frac{1}{T^\pm}\right|.
\end{equation}

The sample impedance is educed by an inverse method introduced in~\cite{auregan2004measurement}. It is based on the multi-modal calculation of the acoustic field in a 2D lined channel with a uniform flow. The Ingard-Myers boundary condition is used to take into account the behaviour of a liner described by its impedance $Z$. The goal of the optimisation procedure is then to find the impedance value that minimises the difference between numerically computed and experimentally measured coefficients of the scattering matrix. As an initial guess, the reactance of a quarter wavelength resonator with a thickness equal to the one of the CD actuator is chosen. The normalised resistance of the initial impedance is taken equal to one. 
More details on the duct facility as well as the measurement techniques used for grazing incidence and grazing flow experiments can be found in\cite{d2020effect}.

\subsection{Results with hybrid absorption}
The choice of the wire mesh used for hybrid absorption technique depends on the target resistance under the grazing incidence.
In a 2D lined duct of height $h$, the impedance $Z_{opt}$ of the infinite wall that maximizes the modal sound absorption is given by the Cremer's formula \cite{tester1973, tester1973_2}: 
\begin{equation}
    Z_{opt} = (0.929-j0.744) \frac{2 f h}{c}.
\end{equation} 
In the present experiment with duct height $h = 40$ mm, the optimal normalized resistance linearly increases from 0.02 to 0.44 in the $100-2000$ Hz frequency range. Thus, for the hybrid absorption method, a wire mesh with a resistance of $0.25$ was chosen to front the CD actuator, since this value falls close to the middle of the target interval. 

\figref{H_M0} (Left) illustrates the transmission losses TL$^+$ and TL$^-$ as a function of the frequency. The amplitude of the incident wave is $90$ dB and the results obtained with and without the hybrid control of the CD actuator are shown. In the passive case, the system consists simply of a cavity covered with the wire mesh, since the corona electrodes are transparent for sound. Thus, sound attenuation is insignificant at frequencies below $600$ Hz and increases towards a peak value of approximately $17$ dB at $1500$ Hz. When the cells are active, sound is absorbed more efficiently at frequencies up to $1550$ Hz. Indeed, the transmission loss achieved with hybrid absorption exceeds 20 dB between 140 and 1500 Hz. For higher frequencies, the advantage of active system becomes negligible and passive attenuation even outperforms the controlled case. One can observe that the transmission losses measured in each direction of acoustic incidence are not totally the same. Such phenomenon occurs due to the need of preliminary calibration of the control system. The algorithm takes into account the propagation paths from the microphones to the actuator estimated prior to the absorption measurement. Since the test duct is not perfectly symmetric relative to the lined section, the obtained filters differ for the two directions of sound propagation, resulting finally in these slightly different transmission curves.

\begin{figure*}[!h]
 \centering
 \includegraphics[width = 8cm]{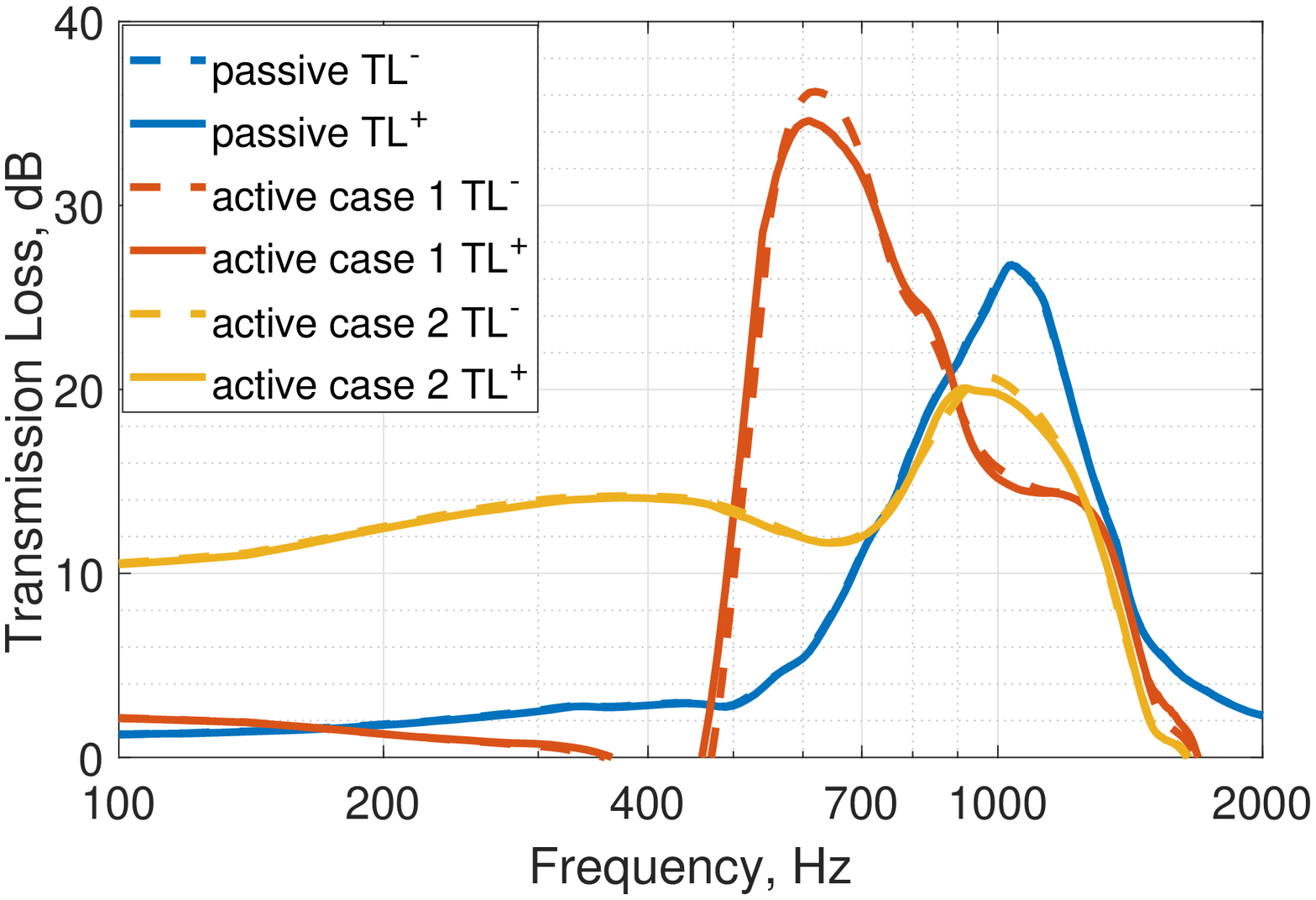}
  \includegraphics[width = 8cm]{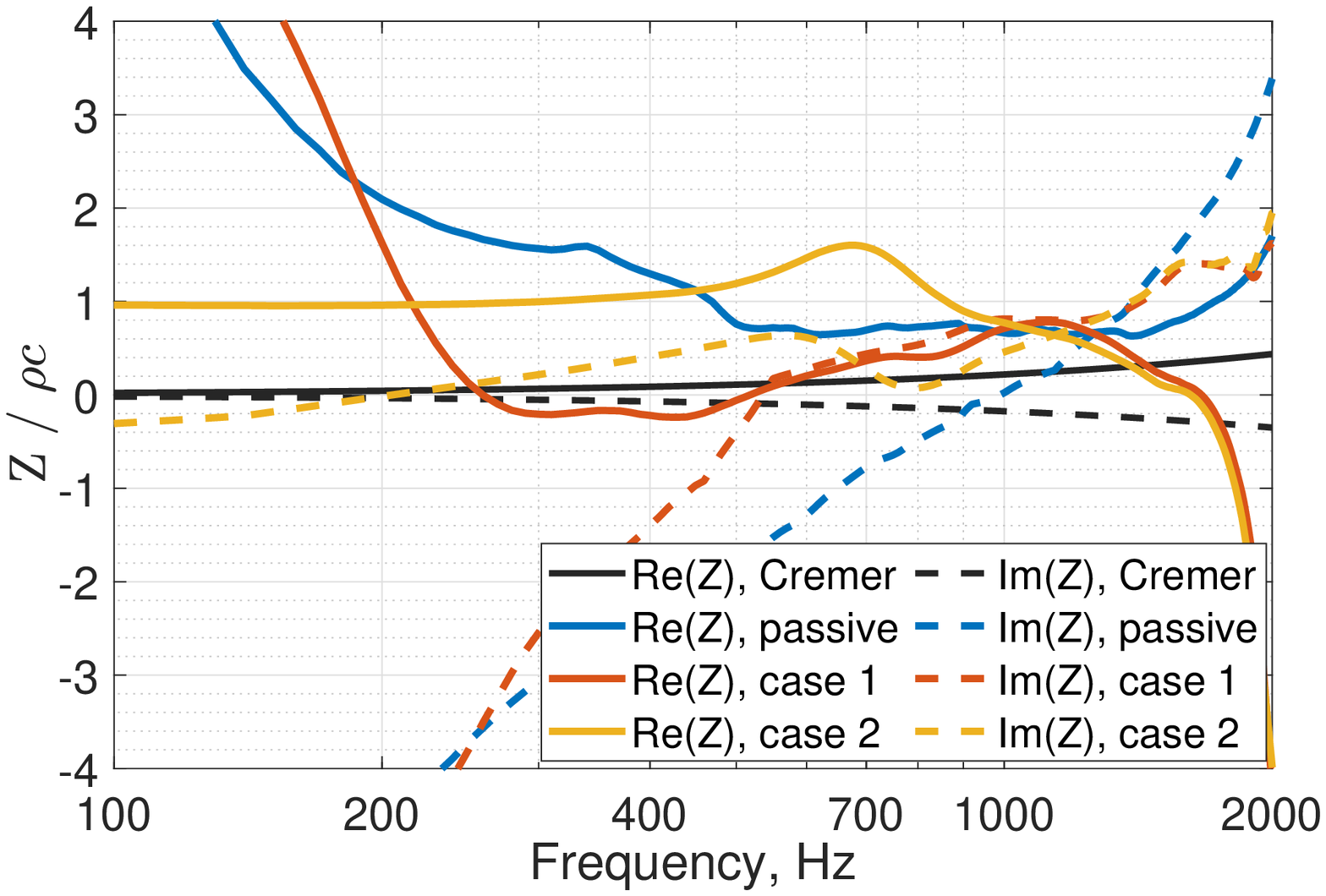}
 \caption{Transmission loss (left) and normalized acoustic impedance (right) of direct impedance control system in comparison to passive behaviour without mean flow. The Cremer impedance is calculated for a duct height of 40 mm.}
 \label{DIC_M0}
\end{figure*}

\figref{H_M0} (right) shows the educed normalized acoustic impedance found using the previously described inverse method. The impedance given by the Cremer formula for $h=40$ mm is also displayed. As can be seen, Cremer's impedance implies both rather low resistance and reactance in the frequency range from $100$ to $2000$ Hz. The passive system impedance is far from these values, except around $1.5$ kHz where the resistance comes to a minimum and the reactance crosses the Cremer optimum. As a consequence, rather high sound absorption is obtained in this region. When the system is controlled, the impedance imposed to the duct wall presents almost constant resistance, around $0.4\rho c$, close to the expected one. The measured reactance evolves in a similar way to the one observed for the normal incidence case. However, its deviation from target zero value starts at lower frequencies, around $200$ Hz. These curves explain the good absorption results displayed in  \figref{H_M0} (left). Indeed, high values of transmission loss are achieved in the low frequency range due to a good match between the target and achieved  reactance, with an overall low resistance. As frequency grows to 1 kHz, the major role in sound absorption transfers to the controlled resistance, as the imaginary part of impedance no longer follows the optimal curve. At higher frequencies, both components of impedance deviate from the Cremer's optimal, leading to the drop in transmission loss in \figref{H_M0} (Left).

\subsection{Results with the direct impedance control}

The direct impedance control method allows targeting a frequency dependent impedance. Moreover, since it does not involve any complex filtering, it can be used without preliminary calibration. For the grazing incidence application the target impedance transfer function $Z_t$ is modelled as follows:

\begin{equation}\label{Zt}
Z_t = s a_m M + a_r R + a_c\frac{1}{s C},
\end{equation}
where $s$ is a Laplace variable. Equation \eqref{Zt} describes the CD actuator backed with a cavity as a sum of an acoustic mass $s M$, a resistance $R$ and a compliance $\frac{1}{s C}$ with their corresponding weights $a_m$, $a_r$, $a_c$. Note that the resistance $R$ has been taken as a constant for a sake of simplicity, although it is not completely true in experiments (see for example \figref{DIC_M0} right, passive case). The values M = 0.044 kg$\cdot$m$^{-2}$, R = 35 Pa$\cdot$s$\cdot$m$^{-1}$ and C = 4.3e-7 m$\cdot$Pa$^{-1}$ were estimated at the front face of the actuator from normal incidence measurements, when the control is off. Thus, when the coefficients $a_m$, $a_r$, and $a_c$ equal 1, this corresponds to the passive case.

The experimental setup is similar to the one depicted in \figref{CD_liner}, except that the part accommodating the control microphone is replaced by a longer section in order to install two sensors. Instead of the wire mesh, a thin layer of Kevlar (weight = $61$ g$\cdot$m$^{-2}$, thickness = $0.12$ mm) is glued on top of the cavity to prevent the flow from entering inside the actuator.

The measurements were first carried out in the absence of mean flow. The transmission losses induced by the active system employed with two different sets of parameters are presented in \figref{DIC_M0} (Left), along with the passive system absorption. Active case 1 corresponds to the setup where $a_m = 0.5$, $a_r = 2$, $a_c = 0.2$, whereas active case 2 targets a purely resistive impedance with $a_m = a_c = 0$ and $a_r = 2$. The passive liner absorbs noise locally around $1050$ Hz. The absorption peak is located at lower frequencies compared to the hybrid absorption case, since the liner cavity is longer. 
When the actuators are active, the frequency of the absorption peak is proportional to $\sqrt{a_c/a_m}$ \cite{rivet2016} and can thus be shifted, either towards the lower or the higher frequency range. The control parameters chosen for the active case 1 (red curve in \figref{DIC_M0} (Left)) decreases the peak absorption frequency. The resulting transmission losses exceed 15 dB from 500 to 1000 Hz. The greatest losses are obtained at $620$ Hz, where the transmission loss reaches approximately $34-36$ dB. Active case 2 (yellow line in \figref{DIC_M0}(Left)) induces generally lower transmission losses, but the sound attenuation remains higher than 10 dB between $100$ and $1500$ Hz. Still, the obtained performance does not reach the one obtained with hybrid absorption. It is worth noting that the final response of the system and the associated noise absorption is the cumulative result of many control parameters such as: input high pass filtering of the signal, target impedance, gains achieved in the feedback loop. Finally, since the target impedance $Z_t$ is controlled in front of the actuator, the achieved impedance at the duct wall is also influenced by the propagation delays to the wall and the additional resistance of the Kevlar layer.

The achieved impedance is depicted in \figref{DIC_M0} (Right) and compared with the Cremer's target. Below 200 Hz, the control is affected by the high pass filtering of the microphone signals. Active case 1 loses passivity/stability at frequencies up to 500 Hz, which is indicated by the negative achieved resistance and negative transmission loss values. Due to reduced $a_m$ and $a_c$ the absolute value of reactance (red dashed line in \figref{DIC_M0}) decreases faster than the passive curve. Together with low resistance such combination leads to a sharp rise of achieved transmission loss. Active case 2 indeed imposes rather constant resistance with relatively low reactance up to 1300 Hz, but the achieved resistance is higher, than with hybrid absorption method (\figref{H_M0}). As a result the sound attenuation is broadband, but presents a lower efficiency than in the case of hybrid absorption. Above approximately 1300 Hz the obtained impedance in both cases deviates from the optimal because of the significant propagation delays compared to the wavelength at this frequency.
Nevertheless, the two very different active behaviours demonstrate the wide scope of possible control cases implementable with direct impedance control. 

\subsection{Mean flow effects}

\begin{figure*}[h]
 \centering
 \includegraphics[width = 8cm]{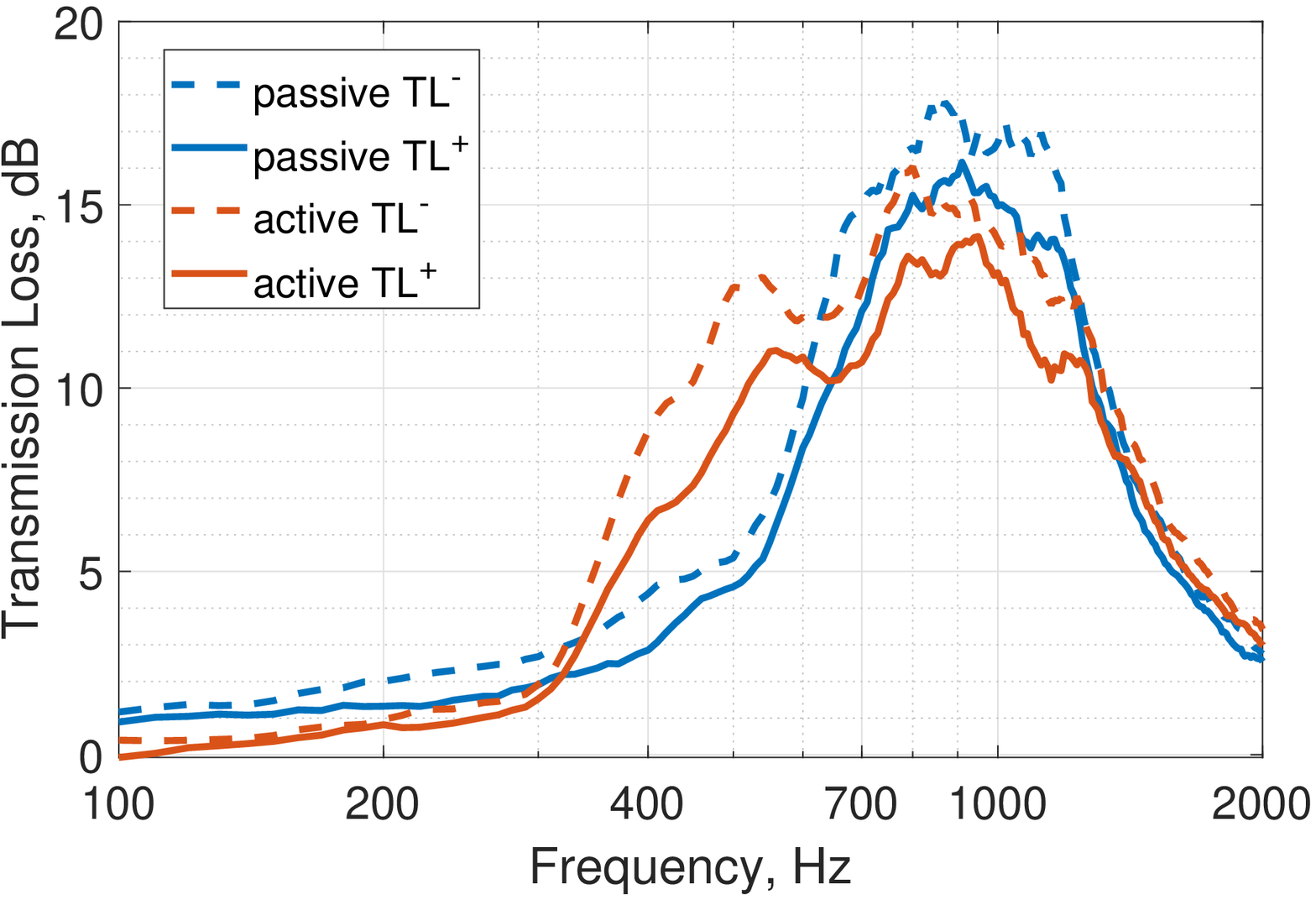}
  \includegraphics[width = 8cm]{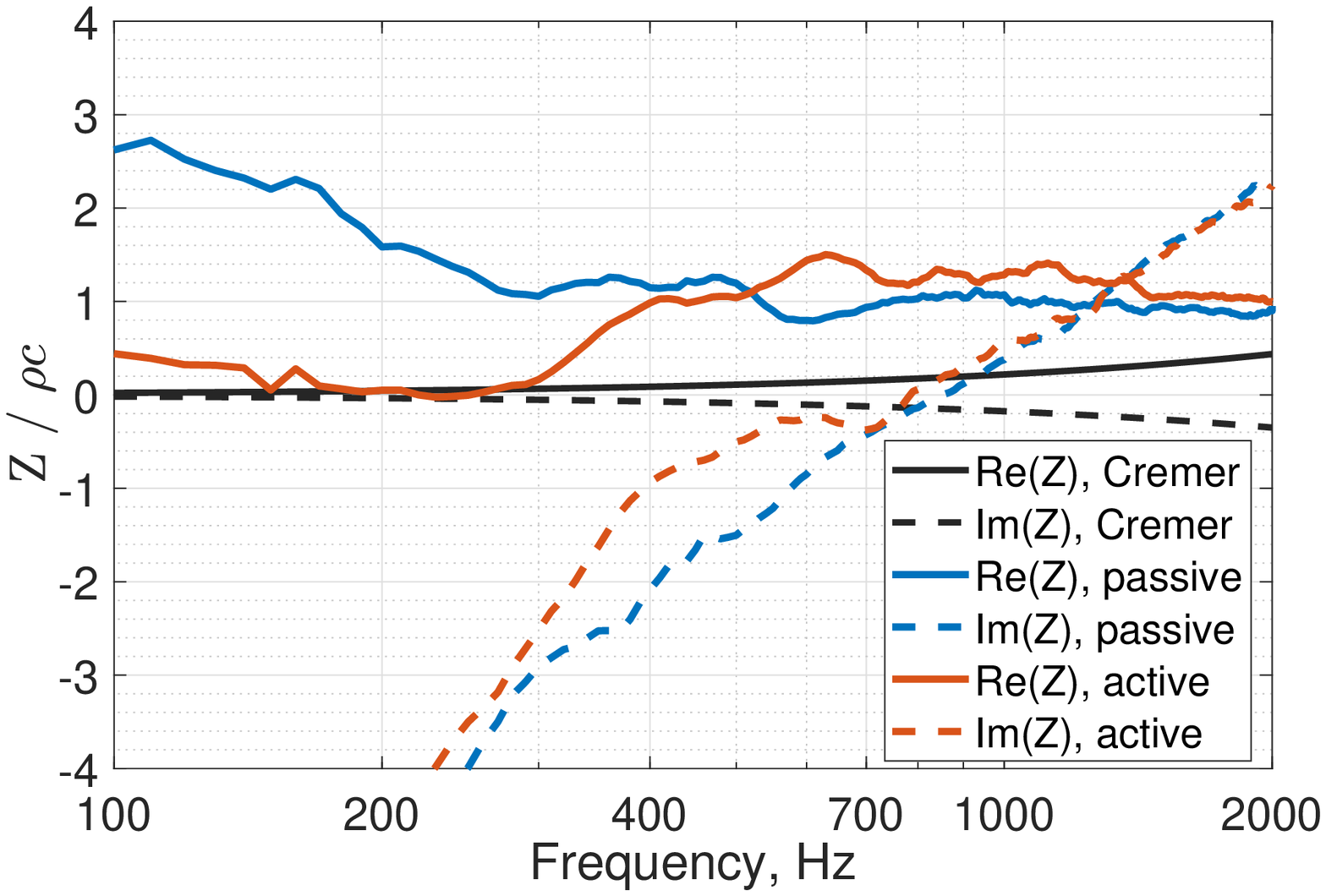}
 \caption{Transmission loss (left) and normalized acoustic impedance (right) of direct impedance control system in comparison to passive behaviour in a mean flow with $M=0.05$. The Cremer impedance is calculated for a duct height of 40 mm.}
 \label{DIC_M005}
\end{figure*}

The direct impedance control strategy is now applied in the presence of a mean flow with a Mach number $M=0.05$ (approximately $15$ m.s$^{-1}$). With one Kevlar layer the output corona signal was saturated due to the flow induced noise. To improve the protection of the liner's microphones from the flow, two layers of Kevlar were mounted on the active cells. As a drawback, the total resistance of the system was increased. The transmission loss achieved with direct impedance control is  presented in \figref{DIC_M005} (Left), along with results for the passive case. For both active and passive cases, the transmission for waves coming from the upstream of the sample is not equal to the one measured for waves coming from the downstream. This loss of reciprocity due to convection effects is classical for flow duct experiments: acoustic waves which propagate with the flow are transmitted better than the ones travelling in the opposite direction. Nevertheless, the active system demonstrates similar performance relative to the passive one. Without control, the liner absorbs sound the most efficiently in the region around 900 Hz. With the direct control of the impedance, transmission loss values are 2-3 dB lower in the same frequency range. Compared to the no flow measurements, weaker maximal absorption is obtained in general. However, the bandwidth of sound attenuation is extended towards the low frequencies

The following control parameters were set for the target impedance: $a_m = 0.5$, $a_r = 2$, $a_c = 0.2$. The corresponding acoustic impedance is depicted in \figref{DIC_M005} (right). Below 200 Hz, the active system does not operate correctly due to the high pass filtering of input signals. In the middle range the reactance approaches zero faster than in the passive case due to the reduced compliance coefficient $a_c$. This explains improved sound absorption. At higher frequencies, the reactance follows closely the passive curve governed by the mass term. Since two layers of Kevlar impose a relatively high resistance, the real part of achieved impedance lays higher than in the experiments without flow. Thus, it yields overall lower transmission losses than in the no flow case. 

One must note that the Cremer's impedance (black curve in \figref{DIC_M005} (right) should be corrected by the factor $1/(1+M)^2$ in the presence of flow \cite{tester1973_2}. Since the flow velocity is reasonably low, the deviation does not exceed 10\% and no change was applied to $Z_t$ for upstream and downstream measurements. For the same reason, the no flow Cremer's impedance is displayed in \figref{DIC_M005} (right). Finally, as in the absence of flow, direct impedance control in experiment with $M=0.05$ allows shifting noise attenuation towards the lower frequency range, compared to the passive case. The tests at higher flow velocities were not carried out successfully, because the signal on the control microphones was considerably distorted by the flow and fan noises. 

Hybrid absorption has also been tested in the presence of flow. Unfortunately, the system is not able to counteract the turbulent flow noise. The sound pressure level at the control microphones was already around $95$ dB without any additional acoustic excitation in the duct. Therefore, the output voltage of the actuator was already saturated and could not completely minimize the acoustic pressure in the cavity behind the wiremesh. Thus, the target impedance could not be reached. Additional work regarding ways to increase the output level and to filter out the turbulent flow noise contribution are needed to tackle the grazing flow problem.


%% file: section_5_conclusion.tex
\label{conclusion}

In this work, a plasma-based electroacoustic actuator generating sound waves by the use of the atmospheric corona discharge has been employed to perform acoustic absorption in real time. To that end, two control strategies have been introduced and tested for both normal and grazing incidence. 

The hybrid absorption method induces high sound absorption in a broadband manner, when an appropriate porous layer is used. In particular, the results for grazing incidence seem to be very competitive with conventional passive technologies in terms of absorption bandwidth and low frequency behaviour. However, since the transducer has a limited acoustic power output\cite{sergeev2020}, the expected performance is only achieved for sound pressure levels up to 90-95 dB. Moreover, only constant and real impedance can be targeted.

Thus, an alternative direct impedance control solution based on pressure-velocity feedback has been proposed. Complex frequency dependent impedance is obtained, and higher incident sound pressure levels are attenuated.  
Moreover, promising results with mean flow are presented in the studied range of velocities. The achieved absorption is extended towards the low frequencies compared to the passive liner. However, the overall absorption values and its frequency bandwidth are lower than with hybrid absorption, mostly due to the way the particle velocity is estimated. Indeed, alternatives should be found in the future to improve or replace the two microphone-based estimation of the particle velocity.

More research should be now addressed to tackle the grazing incidence case with flow. For example, the response of each cell should be taken into account. Since two active cells are operating independently, it is noticed during the measurements, that most of the control effort falls on the cell closest to the noise source, thus partially missing an optimal contribution from the second cell. The control algorithm could be modified in order to better distribute the power among the cells resulting in the increase of the maximal total sound level the active system can sustain. Finally, improvements concerning the modelling and the understanding of the actuator would permit to increase the system output power and to make it tolerate faster flows.

This study revealed several positive aspects of the CD actuators. They offers more flexibility in design and implementation than conventional loudspeakers. In particular, the actuator itself is compact, with a thickness smaller than $1$ cm. Also, its operational area can be easily scaled without any change in frequency response and control voltage ranges if the distance between high voltage wires remains the same (longer actuators could be used if target frequencies to absorb are lower). Similarly, the shape of the actuator can be designed arbitrarily and is not restricted to a rectangular one. All this can bring savings in electronics due to the reduced number of controlled actuators.
 
 Eventually, it appears that the CD technology can be applied for active noise control under normal and grazing incidence and achieve a broadband sound absorption.

%% file: preprint_Sergeev_CD_actuator_as_sound_absorber.bbl
\begin{thebibliography}{10}

\bibitem{palma2018acoustic}
G.~Palma, H.~Mao, L.~Burghignoli, P.~G{\"o}ransson, and U.~Iemma, ``Acoustic
  metamaterials in aeronautics,'' {\em Applied Sciences}, vol.~8, no.~6,
  p.~971, 2018.

\bibitem{ma2020development}
X.~Ma and Z.~Su, ``Development of acoustic liner in aero engine: a review,''
  {\em Science China Technological Sciences}, pp.~1--14, 2020.

\bibitem{maa1998potential}
D.-Y. Maa, ``Potential of microperforated panel absorber,'' {\em the Journal of
  the Acoustical Society of America}, vol.~104, no.~5, pp.~2861--2866, 1998.

\bibitem{beck2015impedance}
B.~S. Beck, N.~H. Schiller, and M.~G. Jones, ``Impedance assessment of a
  dual-resonance acoustic liner,'' {\em Applied Acoustics}, vol.~93,
  pp.~15--22, 2015.

\bibitem{auregan2018ultra}
Y.~Aur{\'e}gan, ``Ultra-thin low frequency perfect sound absorber with high
  ratio of active area,'' {\em Applied Physics Letters}, vol.~113, no.~20,
  p.~201904, 2018.

\bibitem{auregan2016low}
Y.~Aur{\'e}gan, M.~Farooqui, and J.-P. Groby, ``Low frequency sound attenuation
  in a flow duct using a thin slow sound material,'' {\em The Journal of the
  Acoustical Society of America}, vol.~139, no.~5, pp.~EL149--EL153, 2016.

\bibitem{boulvert2019optimally}
J.~Boulvert, T.~Cavalieri, J.~Costa-Baptista, L.~Schwan, V.~Romero-Garc{\'\i}a,
  G.~Gabard, E.~R. Fotsing, A.~Ross, J.~Mardjono, and J.-P. Groby, ``Optimally
  graded porous material for broadband perfect absorption of sound,'' {\em
  Journal of Applied Physics}, vol.~126, no.~17, p.~175101, 2019.

\bibitem{cavalieri2020graded}
T.~Cavalieri, J.~Boulvert, G.~Gabard, V.~Romero-Garc{\'\i}a, M.~Escouflaire,
  J.~Regnard, and J.-P. Groby, ``Graded and anisotropic porous materials for
  broadband and angular maximal acoustic absorption,'' {\em Materials},
  vol.~13, no.~20, p.~4605, 2020.

\bibitem{olson1953electronic}
H.~F. Olson and E.~G. May, ``Electronic sound absorber,'' {\em The Journal of
  the Acoustical Society of America}, vol.~25, no.~6, pp.~1130--1136, 1953.

\bibitem{thenail1994active}
D.~Thenail, M.-A. Galland, and M.~Sunyach, ``Active enhancement of the
  absorbent properties of a porous material,'' {\em Smart Materials and
  Structures}, vol.~3, no.~1, p.~18, 1994.

\bibitem{furstoss1997}
M.~Furstoss, D.~Thenail, and M.-A. Galland, ``Surface impedance control for
  sound absorption: direct and hybrid passive/active strategies,'' {\em Journal
  of sound and vibration}, vol.~203, no.~2, pp.~219--236, 1997.

\bibitem{rivet2016}
E.~Rivet, S.~Karkar, and H.~Lissek, ``Broadband low-frequency electroacoustic
  absorbers through hybrid sensor-/shunt-based impedance control,'' {\em IEEE
  Transactions on Control Systems Technology}, vol.~25, no.~1, pp.~63--72,
  2016.

\bibitem{galland2005}
M.-A. Galland, B.~Mazeaud, and N.~Sellen, ``Hybrid passive/active absorbers for
  flow ducts,'' {\em Applied acoustics}, vol.~66, no.~6, pp.~691--708, 2005.

\bibitem{betgen2012implementation}
B.~Betgen, M.-A. Galland, E.~Piot, and F.~Simon, ``Implementation and
  non-intrusive characterization of a hybrid active--passive liner with grazing
  flow,'' {\em Applied Acoustics}, vol.~73, no.~6-7, pp.~624--638, 2012.

\bibitem{boulandet2018}
R.~Boulandet, H.~Lissek, S.~Karkar, M.~Collet, G.~Matten, M.~Ouisse, and
  M.~Versaevel, ``Duct modes damping through an adjustable electroacoustic
  liner under grazing incidence,'' {\em Journal of Sound and Vibration},
  vol.~426, pp.~19--33, 2018.

\bibitem{rossi1988}
M.~Rossi, {\em Acoustics and electroacoustics}.
\newblock Artech House Publishers, 1988.

\bibitem{moreau2007airflow}
E.~Moreau, ``Airflow control by non-thermal plasma actuators,'' {\em Journal of
  physics D: applied physics}, vol.~40, no.~3, p.~605, 2007.

\bibitem{thomas2008plasma}
F.~O. Thomas, A.~Kozlov, and T.~C. Corke, ``Plasma actuators for cylinder flow
  control and noise reduction,'' {\em AIAA journal}, vol.~46, no.~8,
  pp.~1921--1931, 2008.

\bibitem{el1997drag}
S.~El-Khabiry and G.~Colver, ``Drag reduction by dc corona discharge along an
  electrically conductive flat plate for small reynolds number flow,'' {\em
  Physics of fluids}, vol.~9, no.~3, pp.~587--599, 1997.

\bibitem{kopiev2014instability}
V.~Kopiev, Y.~S. Akishev, I.~Belyaev, N.~Berezhetskaya, V.~Bityurin,
  G.~Faranosov, M.~Grushin, A.~Klimov, V.~Kopiev, I.~Kossyi, {\em et~al.},
  ``Instability wave control in turbulent jet by plasma actuators,'' {\em
  Journal of Physics D: Applied Physics}, vol.~47, no.~50, p.~505201, 2014.

\bibitem{huang2008streamwise}
X.~Huang and X.~Zhang, ``Streamwise and spanwise plasma actuators for
  flow-induced cavity noise control,'' {\em Physics of Fluids}, vol.~20, no.~3,
  p.~037101, 2008.

\bibitem{bastien1987}
F.~Bastien, ``Acoustics and gas discharges: applications to loudspeakers,''
  {\em Journal of Physics D: Applied Physics}, vol.~20, no.~12, p.~1547, 1987.

\bibitem{bequin2007}
P.~B{\'e}quin, K.~Castor, P.~Herzog, and V.~Montembault, ``Modeling plasma
  loudspeakers,'' {\em The Journal of the Acoustical Society of America},
  vol.~121, no.~4, pp.~1960--1970, 2007.

\bibitem{sergeev2020}
S.~Sergeev, H.~Lissek, A.~Howling, I.~Furno, G.~Plyushchev, and P.~Leyland,
  ``Development of a plasma electroacoustic actuator for active noise control
  applications,'' {\em Journal of Physics D: Applied Physics}, vol.~53, no.~49,
  p.~495202, 2020.

\bibitem{zhang2017}
B.~Zhang, J.~He, and Y.~Ji, ``Dependence of the average mobility of ions in air
  with pressure and humidity,'' {\em IEEE Transactions on Dielectrics and
  Electrical Insulation}, vol.~24, no.~2, pp.~923--929, 2017.

\bibitem{klein1954}
S.~Klein, ``Un nouveau transducteur {\'e}lectroacoustique: l'ionophone,'' {\em
  Acta Acustica united with Acustica}, vol.~4, no.~1, pp.~77--79, 1954.

\bibitem{auregan2004measurement}
Y.~Aur{\'e}gan, M.~Leroux, and V.~Pagneux, ``Measurement of liner impedance
  with flow by an inverse method,'' in {\em 10th AIAA/CEAS Aeroacoustics
  Conference}, p.~2838, 2004.

\bibitem{d2020effect}
M.~D'Elia, T.~Humbert, and Y.~Aur{\'e}gan, ``Effect of flow on an array of
  helmholtz resonators: Is kevlar a “magic layer”?,'' {\em The Journal of
  the Acoustical Society of America}, vol.~148, no.~6, pp.~3392--3396, 2020.

\bibitem{tester1973}
B.~Tester, ``The optimization of modal sound attenuation in ducts, in the
  absence of mean flow,'' {\em Journal of Sound and Vibration}, vol.~27, no.~4,
  pp.~477--513, 1973.

\bibitem{tester1973_2}
B.~Tester, ``The propagation and attenuation of sound in lined ducts containing
  uniform or “plug” flow,'' {\em Journal of Sound and Vibration}, vol.~28,
  no.~2, pp.~151--203, 1973.

\end{thebibliography}
